\documentclass{kapproc} 

\RequirePackage{graphicx}%
\RequirePackage{epsf}%
\input{psfig.sty}

\upperandlowercase
\let\footnote\savefootnote
\let\footnotetext\savefootnotetext 
 
\kluwerbib 

\begin{document}


\articletitle{Massive Star Feedback on the IMF}


\chaptitlerunninghead{Massive Star Feedback on the IMF}


\author{M. Robberto\altaffilmark{1}, J. Song\altaffilmark{2}, G. Mora Carrillo\altaffilmark{3}, 
S. V. W. Beckwith, R. B. Makidon, and N. Panagia\altaffilmark{1}}
\affil{Space Telescope Science Institute, 3700 San Martin Drive, Baltimore, MD 21218}
\altaffiltext{1}{Affiliated with the Space Telescope Division of the European Space Agency, ESTEC, Noordwijk, the Netherlands} 
\altaffiltext{2}{Departement of Astronomy, University of Illinois at Urbana-Champaign, USA}
\altaffiltext{3}{Instituto de Astrof\'{\i}sica de Canarias, La Laguna, Tenerife, Spain}


 \begin{abstract}
We have obtained the first measures of the mass accretion rates on stars of the Trapezium Cluster. They appear systematically lower
than those of similar stars in the Taurus-Auriga association. Together with premature disk evaporation, dramatically revealed by
the HST images of the Orion proplyds, this result suggests that low mass stars in a rich cluster 
may be ``dwarfed'' by the influence of nearby OB stars. This feedback mechanism
affects the IMF, producing an excess of low mass stars and brown dwarfs. The observed frequency of low mass objects in Orion vs. Taurus seems to confirm this scenario.
 \end{abstract}

\section{Introduction}
In a recent paper (Robberto et al. 2004), we have presented the first measures of mass accretion rates on stars in
the core of the Orion Nebula Cluster (``Trapezium Cluster''). 
Using the UV excess in the Balmer continuum, measured through the F336W
filter of the {\sl WFPC2} onboard {\sl HST}, and the known correlation between the U-band excess and the 
total accretion luminosity (Gullbring et al. 1998), we estimate accretion rates in the 
range $10^{-8} - 10^{-12}$~M$_\odot$yr$^{-1}$. For stars older than 1~Myr there is some evidence 
of a relation between mass accretion rates and stellar age. Overall, mass accretion rates 
are significantly lower than those measured by other authors in the Orion flanking fields (Rebull et al. 2000) or 
in Taurus-Auriga association (Hartmann et al. 1988). In particular, in the Taurus-Auriga association Hartmann et al. find 
an average accretion rate of $10^{-8}$~M$_\odot$yr$^{-1}$ for a sample containing a similar number of stars with masses and 
ages comparable to ours in the Trapezium cluster. It appears that the main accretion phase in the Trapezium cluster
has been recently terminated, nearly for all sources and independently on their ages.

\section{Disk evolution in a young OB cluster}
Mass accretion is expected to decrease with time and eventually cease with disk exhaustion. In our case,
however, it seems rather artificial to have all disks depleted almost at the same time across such a young cluster ($\sim 1$~Myr).
An attractive possibility is to attribute the drop in mass accretion to a common trigger event caused by an external agent. 
It is known that disks in Orion are photo-ablated by the UV radiation of OB stars, the ``proplyd'' phenomenon 
(O'Dell, Wen, \& Hu 1993, O'Dell \& Wen 1994), with
estimated mass-loss rates $10^{-7}-10^{-6}$~~M$_\odot$yr$^{-1}$ (Churchwell et al. 1987, Henney and O'Dell 1999). When a protostar is exposed to the ionizing flux of a new-born OB star, the disk mass decreases rapidly with time. This may regulate the mass accretion rate through the disk and, 
therefore, to the star. 

\section{Consequences for the IMF}
Hester et al. (1996) have proposed for M16 a scenario in which O-type stars evaporate the dusty cocoons hosting a nearby forming
protostar, suggesting that this process, rather than stellar mass loss, determines the final mass of stars near ionizing sources. In Orion we find evidence that this type of process is active at later phases, when the stars with their circumstellar disks are completely 
exposed to the environmental radiation. 
If low-mass stars terminate their pre-main sequence evolution with masses lower than those they would have reached if disk accretion could have proceeded undisturbed until the final disk consumption,
the IMF of the cluster should be affected. For isolated star formation, i.e. occurring with negligible influence from the environment, the final stellar mass is fixed in the early stages of protostar formation through competitive collapse/fragmentation phenomena regulated by magnetic fields or supersonic turbulence. Most of the 
stellar material is then accreted from the circumstellar disks at a rate that decreases with time. 
There is growing evidence
that low mass stars and sub-stellar mass objects share this type of evolution (Muzerolle et al. 2003), possibly on 
longer timescales. However, if a star is not isolated, envelope and especially disk evaporation will eventually
abort the star formation process. Low mass objects, therefore, remain ``dwarfed"  in a OB cluster,
resulting in a relative overabundance of low mass stars and brown dwarfs.
Within this scenario, one should expect to find systematic differences between
the stellar population in the core of the Orion Nebula and that in other regions, such as the Taurus-Auriga association, 
or the outskirts of the Orion Nebula itself, where the ionizing flux
is negligible. Various authors, in particular Luhmann (2000), have recently reported an overabundance by a factor of 2 in brown dwarfs in Orion relative to Taurus. At the same time, the overabundance of low mass stars and brown dwarfs should be compensated for by a depletion of intermediate mass stars, and it is intriguing to note that Hillenbrand (1997) found a flattening of the IMF in the core of the Trapezium cluster at masses lower than 0.6~$M_\odot$, whereas the overall Orion stellar population follows the Salpeter law down to the completeness limit of her survey, $0.1~M_\odot$. 
This flattening could be the final outcome of the disk depletion, with all low mass stars
cascading into lower mass bins. There is, in conclusion, growing observational evidence of a feedback mechanism intrinsic to the IMF, in the sense that 
the formation and onset of massive stars affects the mass distribution of low mass stars within the same cluster.

%

\end{document}